\documentclass{mem}
\usepackage{txfonts}
\usepackage{balance}
\usepackage{natbib}
\usepackage{graphicx}
\usepackage[a4paper]{hyperref}
\idline{1}{1}

\begin{document}

\title{
Demography of obscured and unobscured AGN: prospects for a Wide Field X-ray Telescope
}

   \subtitle{}

\author{
R. \,Gilli\inst{1} 
\and P. \, Tozzi\inst{2}
\and P. \, Rosati\inst{3}
\and M. \, Paolillo\inst{4}
\and S. \, Borgani\inst{2,5}
\and M. \, Brusa\inst{6}
\and A. \, Comastri\inst{1}
\and E. \, Lusso\inst{7}
\and F. \, Marulli\inst{7}
\and C. \, Vignali\inst{7}
\and the WFXT team
         }

  \offprints{R. Gilli}

\institute{
INAF --
Osservatorio Astronomico di Bologna, Via Ranzani 1,
I-40127 Bologna, Italy \email{roberto.gilli@oabo.inaf.it}
\and
INAF --
Osservatorio Astronomico di Trieste, Via Tiepolo 11,
I-34131 Trieste, Italy
\and
ESO, Karl Schwarzschild Strasse 2, 85748 Garching bei Muenchen, Germany
\and
Universit\`a Federico II, Dip. di Scienze Fisiche C.U. Monte S.Angelo, ed.6, via Cintia 
80126, Napoli
\and
Dip. di Astronomia dell'Universit\`a di Trieste, Via G.B. Tiepolo 11, I-34131 Trieste, Italy
\and
MPE, Giessenbachstrasse, 1, 85748 Garching bei Muenchen, Germany
\and
Dip. di Astronomia, Universit\`a di Bologna, Via Ranzani 1, 40127 Bologna, Italy
}

\authorrunning{Gilli }

\titlerunning{AGN demography}

\abstract{We discuss some of the main open issues in the evolution of Active Galactic Nuclei
which can be solved by the sensitive, wide area surveys to be performed by the proposed Wide Field X-ray Telescope
mission. 

}
\maketitle{}

\section{Introduction}

The observed scaling relations between the structural properties of massive galaxies (bulge mass, luminosity,
and stellar velocity dispersion) and the mass of the black holes (BHs) at their centers suggest that
galaxy assembly and black hole growth are closely related phenomena. This is often referred to as 
BH/galaxy co-evolution. The BH growth is thought to happen primarily through efficient accretion phases accompanied by the release of kinetic and 
radiative energy, part of which can be deposited into the galaxy interstellar medium.
Active Galactic Nuclei (AGN) are then believed to represent a key phase across a galaxy's lifetime. Support for this hypothesis comes from several lines of evidence such as i) the match between the mass function of BHs grown through AGN phases and that observed in local galaxies \citep{marconi04,shankar04}; ii) the cosmological ``downsizing'' of both nuclear activity and star formation \citep{ueda03,cowie96};  iii) the feedback produced by the AGN onto the galaxy interstellar medium through giant molecular outflows \citep{feruglio10}.
A number of semi-analytic models (SAMs) have been proposed over about the past decade to explain the BH/galaxy co-evolution \citep{kh00,pigi07,menci08,marulli08,lamastra10}. 
These models follow the evolution and growth of dark matter structures across cosmic time, either through the Press-Schechter formalism or through N-body simulations, and use analytic recipes to treat the baryon physics within the dark matter halos. A common assumption of these models is that mergers between gas-rich galaxies trigger nuclear activity and star formation. Recently, a BH/galaxy evolutionary sequence associated to ``wet'' galaxy mergers has been proposed \citep{hop08}, in which an initial 
phase of vigorous star formation and obscured, possibly Eddington limited, accretion is followed by a phase
in which the nucleus first gets rid of the obscuring gas shining as an unobscured QSO, then quenches star formation, and eventually 
fades, leaving a passively evolving galaxy. 

While being successful in many respects, this picture is still very partial and many fundamental pieces
are missing to get a satisfactory understanding of BH/galaxy co-evolution, such as the very first stages of this joint evolution (e.g. at redshifts $z>6$), the cosmological evolution of nuclear obscuration, the triggering mechanisms and environmental effects of nuclear activity.
In this contribution it will be shown how the proposed Wide Field X-ray Telescope mission (WFXT) can
effectively address some of these fundamental issues through sensitive, large area X-ray surveys.

\section{The WFXT surveys at a glance}

As detailed in Rosati et al. (2010, this volume), WFXT is a mission thought and designed to perform X-ray surveys, featuring a large (1 deg$^2$) field-of-view (FOV), large (1 m$^2$ at 1 keV) effective area, and sharp (5'' HEW) resolution, constant across the FOV (the quoted values refer to the goal mission design). 
The observational survey strategy with WFXT will consist of three main X-ray surveys with different area and depth to sample objects
in a wide range of  redshifts and luminosities. To illustrate
the power of the WFXT surveys, a simple excercise can be performed by rescaling the number of X-ray
sources obtained from well-known X-ray surveys of similar sensitivity.
The 20000 deg$^2$ WFXT-Wide survey will cover about 2000 times the area of the XBootes
survey \citep{murray05}, in which more than 3000 X-ray sources have been detected, with the same depth. Similarly,
the 3000 deg$^2$ WFXT-Medium and 100 deg$^2$ WFXT-Deep surveys will cover 3000 times the area of the C-COSMOS
survey \citep{elvis09} and 1000 times the area of the 2Ms CDFS survey \citep{luo08}, in which 1700 and 450 objects have been detected, respectively. By summing these numbers it is easy to see that the
total source sample obtained by WFXT, mainly composed by AGN,  will contain more than
10 million objects.

A more precise estimate of the number of AGN to be detected is obtained by
considering the logN-logS relationships in the soft, 0.5-2 keV, and hard, 2-7 keV, bands.
About 15 millions AGN detections are expected in the soft band up to $z>6$ and about 4 millions
in the hard band. Remarkably, for a  very large number of objects it would be possible to obtain an accurate
spectral characterization over the 0.5-7 keV WFXT band pass and derive physical properties such as the X-ray absorption
(see Section 4 and Matt \& Bianchi, 2010, this volume). Indeed, it is worth noting that, while being most effective at 1 keV, 
the WFXT collecting area at $>4$ keV is still significantly large: in the goal design, the effective area at 5 keV is 
equal to that of XMM (pn+2MOS combined) at the same energy. 

In the following Sections we will present a few unique science cases which can be addressed only with 
wide area and sensitive X-ray surveys.

\section{Supermassive Black Holes in the early Universe}

Most stars formed at $0.5<z<3$, when SMBHs were also growing most of their mass,  
but the assembly of the first organized structures started at earlier epochs, as soon as baryons were able to cool within dark
matter halos. Discovering the first galaxies and black holes ever formed and understanding how their (concurrent?) formation takes 
place is a fascinating subject to which more and more efforts will be devoted in the next decade.
To date, several tens of galaxies and about 20 QSOs have been confirmed spectroscopically at redshifts above 6 \citep[e.g.][]{tani08,fan06,willott09}.
Besides a GRB discovered at $z\sim8.2$ \citep{salvaterra09,tanvir09} and a galaxy at $z=8.6$ \citep{lehnert10}, no other object is known spectroscopically 
at $z>7$, although more than 100 galaxy candidates have been recently isolated through deep near-IR imaging with WFC3 \citep{oesch10,wilkins10}. The most distant SMBHs known to date are three QSOs at $z\sim6.4$ discovered through the SDSS \citep{fan06} and the CFHQS survey \citep{willott09,willott10}. These two surveys
have been the main resource for investigating the most distant QSOs in the past decade.
About 20 QSOs at $z>5.7$ have been detected in the SDSS main survey, which represent the brightest tail of the early QSO population
(log$L_x\sim45.5$; log$L_{bol}\sim47$). Less luminous objects have been detected in the SDSS deep stripe \citep{jiang09} and CFHQS. The BH masses measured for the SDSS bright objects are of the order of $10^9\;M_{\odot}$ \citep{kurk07, kurk09}, which must have been built in less than 1 Gyr, i.e. the age of the Universe at $z=6$. These giant black holes are thought to have formed through mass accretion onto smaller seeds with mass ranging from $10^2M_{\odot}$, 
as proposed for the remnants of massive, PopIII stars \citep{mr01}, to $10^4M_{\odot}$, as proposed for the products of direct collapse of large molecular clouds \citep{volonteri08}. Whatever the seed origin is,  to account for the observed BH mass distribution at $z\sim6$, BH growth must have proceeded almost continuously at (or even above) the Eddington limit for a time interval  of about 1 Gyr. Recent hydrodynamical simulations have shown that, within large dark matter halos, merging between proto-galaxies at $z\sim14$ with seed BH masses of $10^4M_{\odot}$ may trigger Eddington limited nuclear activity to produce a $10^9$ solar mass BH by $z\sim6$ \citep{li07}. The overall frequency and efficiency by which this merging and fueling mechanism can work is however unknown.

Because of the paucity of observational constraints, the study of BH and galaxy formation at early epochs remained speculative. For example, the relation between the BH mass and the galaxy gas mass has been measured for a few bright QSOs at $z\sim5-6$, in which the ratio between the BH mass and the dynamical mass within a few kpc radius, as measured from CO line observations, is of the order of 0.02-0.1 \citep{walter04}. This is more than one order of magnitude larger than the ratio between the BH mass and stellar bulge mass measured in local galaxies. Despite the uncertainties on these measurements (see e.g. \citealt{nara08}), the possibility that SMBHs are leading the formation of the bulge in proto-galaxies, poses severe constraints to galaxy formation models. Large statistical samples of $z>6$ AGN are needed to understand the relation between the BH growth and formation of stars in galaxies at their birth. Based on SAMs (see e.g. \citealt{hop08} and \citealt{marulli08} for recent works), the luminosity function and spatial distribution (clustering) of AGN at $z>6$ would constrain: i) where, i.e. in which dark matter halos, they form; ii) what their average accretion rate is; iii) what the triggering mechanism (e.g. galaxy interactions vs secular processes) is.

\begin{figure}[t!]
\resizebox{\hsize}{!}{
\includegraphics{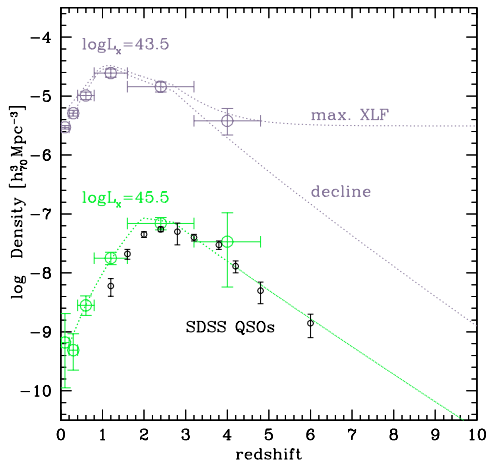}}
\caption{\footnotesize The space density of unabsorbed AGN with log$L_x=43.5$ and log$L_x=45.5$. Big open circles are from \citet{hms05}. 
Small black symbols have been obtained from the space density of optically luminous SDSS QSOs \citep{richards06} by assuming $\alpha_{ox}=-1.5$. The space density of luminous QSOs is found to decrease from $z\sim3$ to $z\sim6$ and is well fitted
by an exponential decline curve (dotted line from \citealt{gch07}). The space density of lower luminosity objects is unconstrained for $z>4$, and we
show two alternative working hypotheses: either a ``decline'' scenario, as observed for luminous QSOs, or a constant density model, which is referred to as ``maximum XLF''.
}
\label{xvol}
\end{figure}

The accreting SMBHs at $z>6$ known to date have been all selected as $i$-band dropouts ($i-z\gtrsim2$) in optical surveys. The sky density of the bright ($z_{AB}<20.2$) QSOs at $z>6$ in the SDSS main survey is 1/470 deg$^2$. Fainter objects ($z_{AB}\sim21-22$) have densities of 1 every $\sim$30-40 deg$^2$ \citep{jiang09,willott09}. The discovery of significant samples of objects therefore relies on the large areas covered by these surveys ($>$8000 deg$^2$ for the SDSS, $\sim$900 deg$^2$ for the CFHQS). Based on the SDSS and CFHQS samples, a few attempts to measure the luminosity function of $z\sim6$ QSOs have been performed, the most recent of which are those by \citet{jiang09} and \citet{willott10}. While the uncertainties at low luminosities 
($M_{1450}>-25$) are still substantial, the space density of luminous QSOs ($M_{1450}<-26.5$; $L_{bol}>10^{47}$ erg s$^{-1}$) is relatively well constrained, decreasing exponentially from $z\sim3$ to $z\sim6$ (see Fig.~\ref{xvol}).
Furthermore, multiwavelength studies of SDSS QSOs at $z\sim6$ showed that these objects have metallicities and spectral energy distributions pretty similar to those of lower redshifts QSOs, suggesting their are already “mature” objects in a young Universe (see however \citealt{jiang10}
for two counter-examples).

Most of the multiwavelength studies mentioned above refer to optical QSOs above the knee of the luminosity function. These likely represent a minor fraction of the active BH population at $z>6$, which is expected to be made primarily by less massive, $\approx10^6\;M_{\odot}$, and less luminous, $10^{44}$ erg s$^{-1}$, objects.

When compared with optically selected objects at the same redshift, X-ray selected AGN are on average less luminous, either intrinsically or because of obscuration effects. Therefore X-ray selection might be the key to sample the bulk of the high-z AGN population.
Several arguments indeed suggest that obscured AGN should be abundant even at very high redshift. First, observations and modeling showed that obscured AGN outnumber unobscured ones by a significant factor up to $z\lesssim4$ and it is then reasonable to extrapolate that even at $z>6$. Second, most current models of galaxy formation postulate that the early phases of accretion onto seed black holes are obscured \citep[e.g.][]{menci08}. Third, optical/near-IR spectroscopy and IR/sub-mm imaging of $z\sim6$ QSOs showed that dust and metals are abundant in their inner regions \citep{juarez09,beelen06}. Dust and metals must have then formed in large quantities at $z>6$ and can effectively absorb the nuclear radiation from the optical regime to the soft X-rays. 

\begin{figure*}[t!]
\resizebox{\hsize}{!}{
\includegraphics{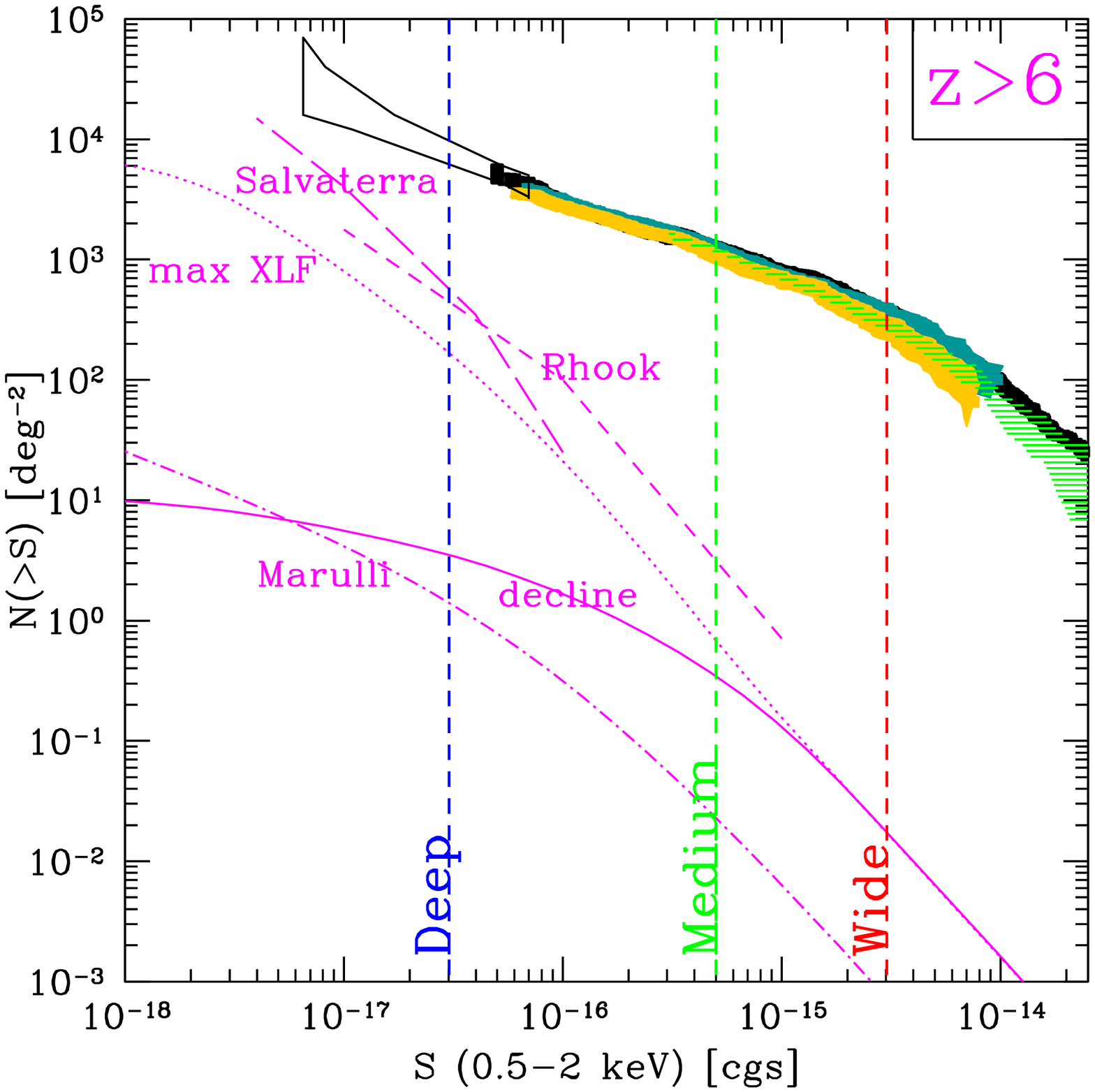}
\includegraphics{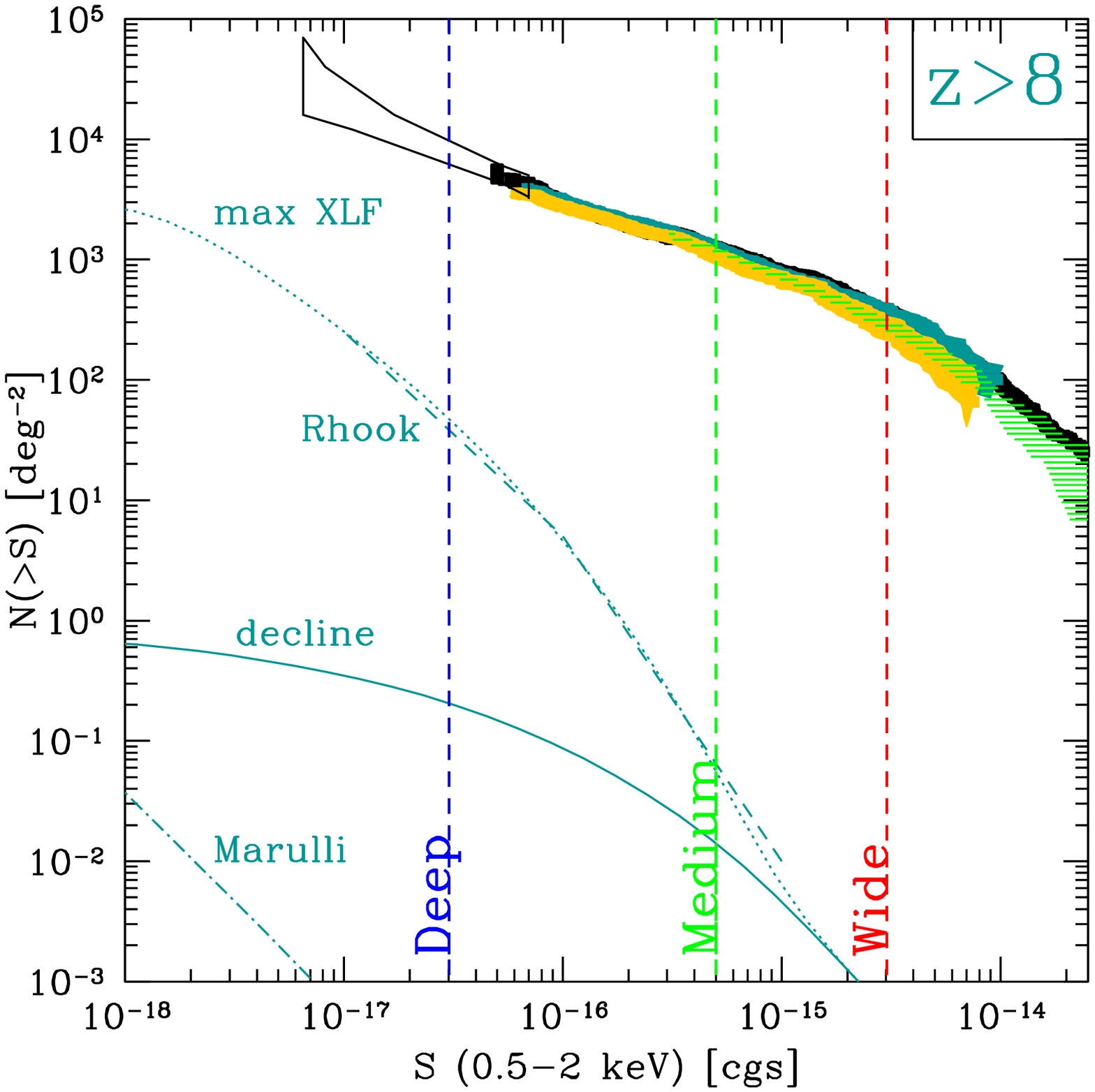}}
\caption{\footnotesize {\it Left:} Expected number counts in the 0.5-2 keV band for $z>6$ AGN according to different models as labeled (see text for details). The datapoints and shaded regions show the observed, total logN-logS of X-ray sources. The sensitivity limits of the three
WFXT surveys are shown as vertical dashed lines. {\it Right}: As in the {\it left} panel but for $z>8$ AGN.
}
\label{lnls}
\end{figure*}

Among the objects discovered so far at $z>6$ no one is obscured, since these have been selected only through optical color selection criteria.
The most distant object discovered to date through X-ray selection is an AGN at $z=5.4$  in  the COSMOS field (Civano et al. in preparation) and only a few are known at $z>4$. The lack of sufficient sky area covered to deep sensitivities is the reason behind that. The discovery space for early obscured AGN is therefore huge, and can have a significant impact on our understanding of BH and galaxy formation.

\begin{figure*}[t!]
\resizebox{\hsize}{!}{
\includegraphics[angle=270]{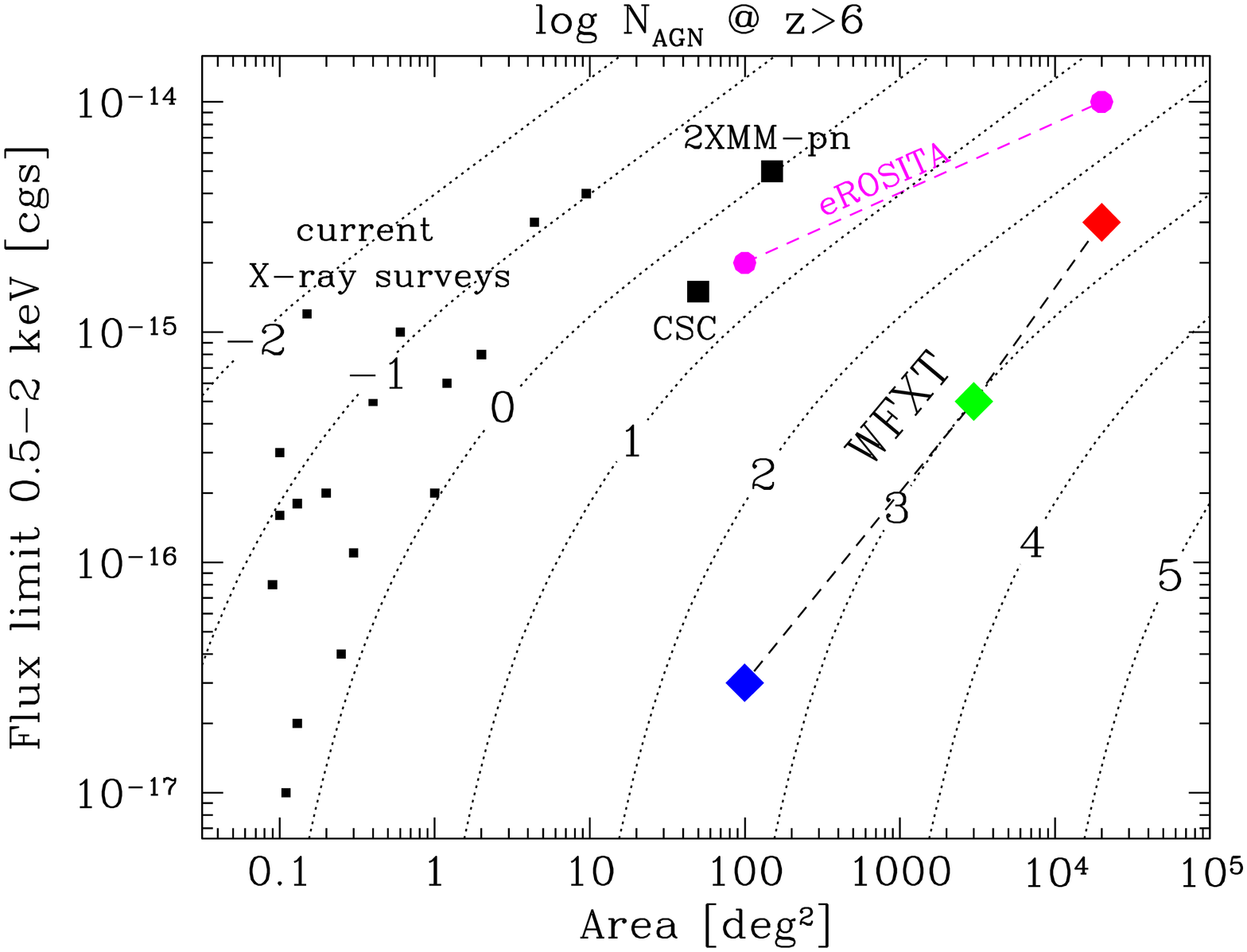}}
\caption{\footnotesize {\it Left:} Number of AGN at $z>6$ expected from the ``decline'' model (see text) for different combination of survey area vs 0.5-2 keV limiting flux.
Dotted lines are the locii of equal AGN number as labeled (labels are in log units). 
Small black squares show the major Chandra and XMM surveys (either performed or ongoing, including the 4Ms CDFS, the 2Ms CDFN, C-COSMOS, AEGIS, XBootes and many
others). Big black squares show the current coverage of the Chandra (CSC) and XMM (2XMM-pn) archives. The deep and wide surveys to be performed with eROSITA in 4 years are shown as magenta circles (based on Cappelluti et al. 2010, this volume). The three WFXT surveys (5 year program) are shown as rotated squares. 
Less than one $z>6$ AGN is expected in each of the current X-ray surveys. eROSITA will return about 40 objects with the current design. WFXT will detect about 1600 objects. 
}
\label{ison}
\end{figure*}

Different routes can be tried to estimate the space and surface density of X-ray emitting high-z AGN and hence make forecasts for the samples
to be observed in the WFXT surveys. As a first step, we considered extrapolations based on our current knowledge of the AGN
X-ray luminosity function (XLF) and evolution. As shown in Fig.~\ref{xvol}, the space density of luminous, unobscured QSOs with $M_{1450}<-26.5$
has been  determined to steadily decrease from $z\sim3$ to $z\sim6$ \citep{richards06}. When their optical luminosity is
converted to the X-rays (using $\alpha_{ox}=-1.5$), there is a good match between their space density
and that determined by \citet{hms05} for soft X-ray selected QSOs in the overlapping redshift range. The behaviour of lower luminosity sources, log$L_x=43.5$, is instead known only 
up to $z\sim 4$ \citep{yencho09,aird10} and extrapolations to higher redshifts are highly uncertain (see Brusa et al. 2010, this volume).
Starting from the X-ray background synthesis model by \citet{gch07}, we considered a first scenario 
in which the space density of AGN with log$L_x<44$ undergoes the same exponential decline as observed for luminous
QSOs - which we will call the ``decline'' scenario - and a second, perhaps extreme, scenario in which it stays
constant above a redshift of $\sim 4$, which will be referred to as the ``maximum XLF'' scenario (see Fig.~1).
The 0.5-2 keV logN-logS of $z>6$ AGN expected in the two above mentioned scenarios is shown in Fig.~\ref{lnls}, where we also show
a range of predictions based on different SAMs of BH/galaxy coevolution \citep{salvaterra07, rh08, marulli08}.
The limiting fluxes of the three WFXT surveys are also shown.
It is clear how the $z>6$ AGN Universe is a completely uncharted territory, the various predictions differing by a few orders of magnitude already at fluxes
around $10^{-16}$ cgs. The predictions by the SAMs of \citet{salvaterra07} and \citet{rh08} are even more optimistic than the ``maximum XLF'' scenario, predicting about 500-600 $z>6$ AGN deg$^{-2}$ at the WFXT-Deep limiting flux. The model by \citet{marulli08} is 
instead the most pessimitic, with 15 $z>6$ AGN deg$^{-2}$ at the same limiting flux. The predictions by SAMs depend on a pretty large number of parameters such as the mass of seed BHs, their location in the density field (i.e. the abundance of the dark matter halos hosting them), the recipes used for accretion, the AGN lightcurve. Although different assumptions are made by different models, the $\sim 3$ orders-of-magnitude difference at faint X-ray fluxes between the predictions by \citet{marulli08} on the one hand, and \citet{salvaterra07} and \citet{rh08} on the other hand, seems to be primarily related to the assumed space density of seed black holes. In the \citet{marulli08} model, seed BHs are placed in each halo that can be resolved by the Millennium simulation, i.e. in those halos with mass above $\sim 10^{10}\;M_{\odot}$, while in \citet{salvaterra07} BHs populate mini-halos with mass as low as 
$\sim 10^{7-8}\;M_{\odot}$, which are therefore much more abundant. In \citet{rh08}  BHs are assumed to radiate at the Eddington limit and their mass to be proportional to that of the hosting halos. This implies that, if small mass seed are assumed, these populate low mass, abundant halos. If large mass seeds are assumed, these populate less abundant halos but their luminosity, and hence detectability, is higher. In each case, surface densities as high as 500 deg$^{-2}$ at the WFXT-Deep limiting flux are reached.

\begin{table}
\caption{Number and minimum 0.5-2 keV luminosity of high-z AGN expected in the WFXT surveys according to the two evolution models (``decline'' and ``max LF'') described in the text.}
\begin{tabular}{lrrr}
\hline \hline
 Quantity& & Survey&  \\
\hline
& Deep& Medium& Wide\\ 
\hline
$z>6$& & & \\
\hline
log$L^{min}$(0.5-2 keV)& 43.1& 44.3& 45.1\\
N. AGN (decline)& 300& 1000& 300\\
N. AGN (max LF)& 15000& 2300& 300\\
\hline
$z>8$& & & \\
\hline
log$L^{min}$(0.5-2 keV)& 43.4& 44.6& 45.4\\
N. AGN (decline)& 20& 45& 10\\
N. AGN (max LF)& 4300& 210& 10\\
\hline                                             
\end{tabular}
\label{bop}
\end{table}

Observations of significant samples at $z>6$ would constrain the physics of early BH formation disentangling between these scenarios. The surveys performed by WFXT have the power to provide such large statistical 
samples. As a reference model, we will consider the ``decline'' model outlined above, which is found to be in excellent agreement with the
0.5-2 keV logN-logS of AGN at $z>3,4,5$ \citep[Civano et al. in prep.]{brusa09}. About 1600 AGN at $z>6$ are expected to be detected
by WFXT, and about 70 ot them should be at $z>8$. A summary of the predictions for AGN at $z>6$ and $z>8$ in the three WFXT surveys
for the decline and maximum XLF scenarios is shown in Table~1. In the WFXT-Deep survey it will be possible to detect  in significant numbers $z>6$ AGN down to log$L_x=43.1$ and $z>8$ AGN 
down to log$L_x=43.4$, making possible to determine the shape of the high-z XLF.
To compare the high-z AGN yields from WFXT surveys with what is expected from major current X-ray surveys and with other X-ray missions which are either proposed or planned, we show in Fig.~\ref{ison} the number of $z>6$ AGN to be detected with different combination of surveyed sky area and soft limiting flux according to the ``decline'' scenario. 
It is evident that even the major current X-ray surveys (e.g. CDFS, CDFN, COSMOS, AEGIS, X-Bootes) should return 
less than one object per field. In the full Chandra and XMM archives (labeled as CSC and 2XMM-pn, respectively) one would expect less than 10 objects in total.
With the current mission design and survey strategy (Cappelluti et al. 2010, this volume), the eROSITA satellite, planned for launch in 2012, is expected to return 
about 40 $z>6$ AGN, most of them at the luminosities of bright SDSS QSOs. The big leap is clearly expected to be performed by WFXT. The mission design and
observing strategy studied for the Internation X-ray Observatory (IXO) are still uncertain: IXO will probably return a sample of $z>6$ AGN with size in between those by eROSITA and WFXT (see Comastri et al. 2010, this volume).

The identification of these objects will clearly require wide area surveys with deep multiband optical and near-IR imaging like e.g. the LSST surveys (Brusa et al. 2010, this volume). On the other hand, the WFXT surveys will represent a perfect complement for all optical and near-IR campaigns that search the unobscured part of the high-z AGN population as image dropouts. 
Late type brown dwarfs are the main contaminants in optically selected samples, being 15 times more abundant than high-z QSOs of comparable magnitudes \citep{fan01}. 
Even when using near-IR colors to separate them from high-z QSO candidates, the success rate of optical spectroscopy is only 20-30\% \citep{jiang09}. If sensitive X-ray images were available underneath each optical dropout, a detection in the X-rays would almost automatically ensure that the object is an AGN, since brown dwarfs of comparable optical magnitudes are $\sim 300$ times fainter in the X-rays. LSST and WFXT surveys will then reinforce each other in the search of $z>6$ AGN.

\begin{figure*}[t!]
\resizebox{\hsize}{!}{
\includegraphics{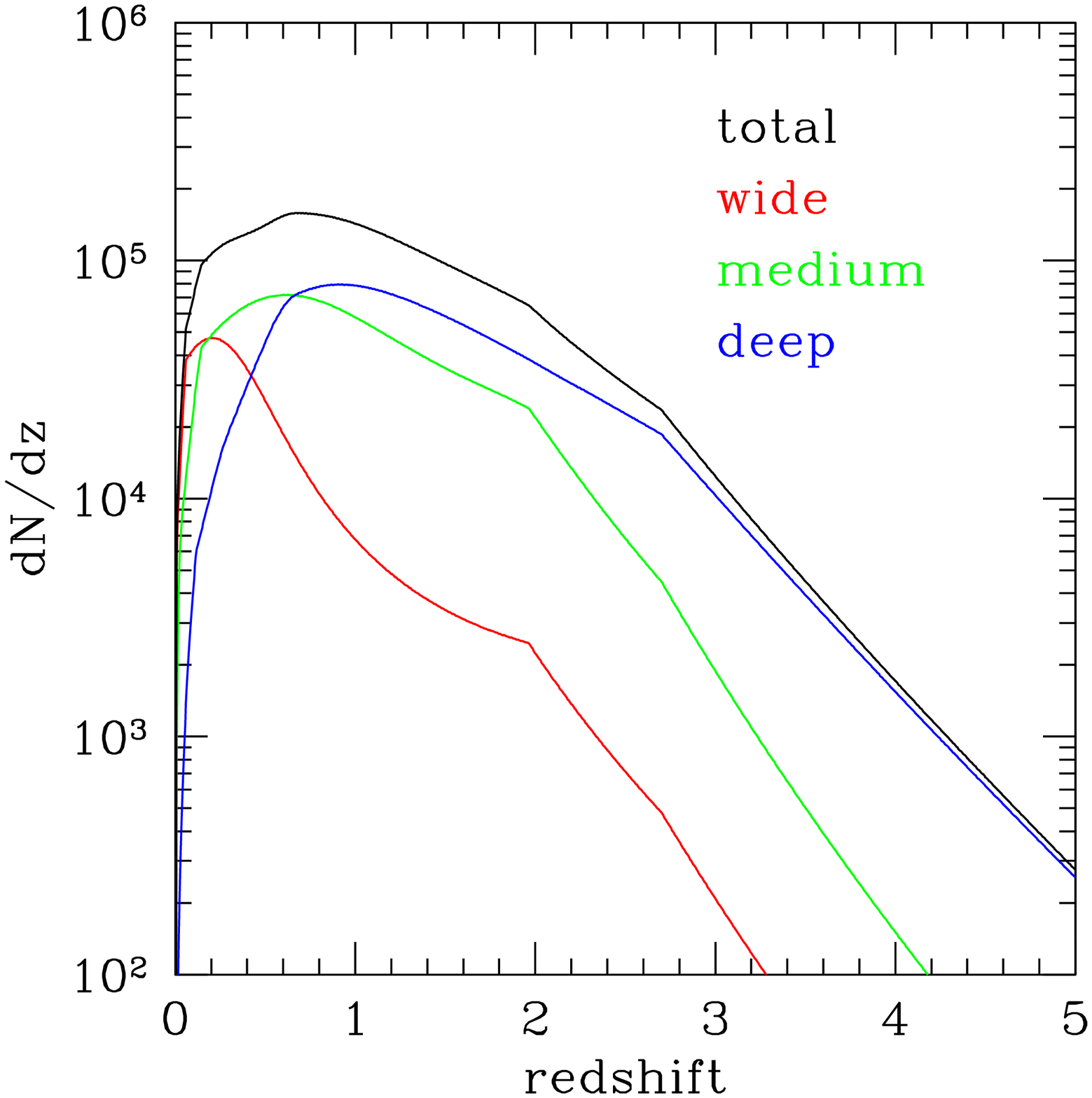}
\includegraphics{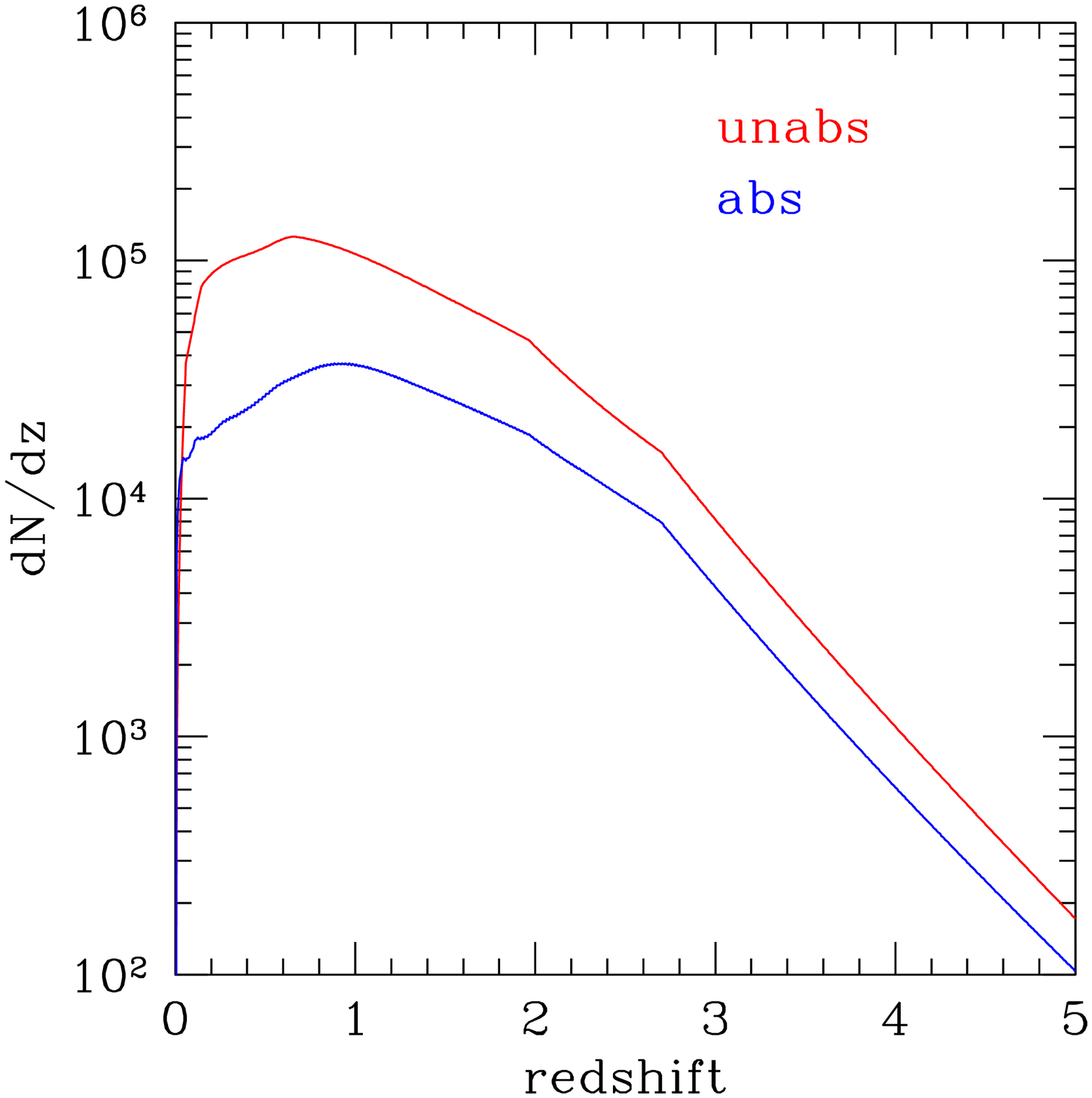}}
\caption{\footnotesize {\it Left}: Redshift distribution of the 300,000 AGN with good X-ray spectra (i.e. with more than 1000 photons) 
in the three WFXT surveys. The Deep survey is providing most of the good X-ray spectra at $z>1$. {\it Right}: As in the {\it left} panel
but splitting the total sample of 300,000 AGN into unabsorbed (log$N_H<22$) and absorbed (log$N_H\geq22$) objects.
}
\label{zdist}
\end{figure*}

\section{Evolution of the obscuration}

A built-in feature of the BH/galaxy evolutionary sequence described in the Introduction is that an obscured accretion phase preceeds a clean accretion phase, at least in powerful, QSO-like objects. One may therefore wonder whether the fraction of obscured AGN was higher in the past. This indeed depends on many parameters such as i) the physical scale of the absorbing gas and how this is driven towards the BH; ii) the relative timescales of the obscured and unobscured phases; iii) whether the absorbed-to-unabsorbed AGN transition occurs also in low-mass/low-luminosity objects (i.e. Seyfert galaxies). Not many theoretical predictions on the evolution of the obscured AGN fraction are available in the literature. Models that relate the obscuration on nuclear scales to the availability of gas in the host galaxy generally predict  an increase of the obscured AGN fraction with redshift \citep[e.g][]{menci08} since the gas mass in galaxies was larger in the past. Some others models, which anti-correlate the covering factor of the obscuring medium to the BH mass and then follow the evolution of the BH mass function using empirical relations \citep{lamastra08}, however do not predict such an increase. The situation is also debated from an observational point of view. An increase of the obscured AGN fraction
with redshift among X-ray selected AGN has been observed by \citet{lafranca05,tu06,h08,trump09}, but other works did not find any evidence of this trend \citep{ueda03,dp06,gch07}. Selection effects in these computations are very important and could mimic a spurious evolution of the obscured AGN fraction (see e.g. \citealt{xray2009}).

Much of this uncertainty can be related to the lack of large statistical samples of AGN with high-quality X-ray spectra. 
Using samples of limited size, which cannot be split into narrow luminosity and redshift bins, it is often complicated to disentangle any 
redshift-dependent from luminosity-dependent effects (it is generally agreed that the obscured AGN fraction decreases with luminosity, 
but the slope and normalization of this relation differ from paper to paper). Furthermore, the absorption column density is often estimated either from
the X-ray hardness ratio, or from the absence of broad emission lines in the optical spectrum. Both methods suffer from caveats and limitations \citep{brusa10}, and ideally the absorbing column should be measured through X-ray spectra with good photon statistics. The most recent XLF determinations contain $\lesssim 2000$ AGN \citep{h08,silverman_xlf,yencho09,ebrero09,aird10}, most of which are just slightly more than simple X-ray detections. With WFXT one would expect to detect about 300,000 AGN with good X-ray spectra, i.e. with more than 1000 photons each, in a broad redshift range ($z\sim0-5$). As shown in Fig.~\ref{zdist}~{\it left}, the WFXT-Deep survey is providing most of the good X-ray spectra at $z>1$, while the Medium and Wide surveys provide the largest contribution at lower redshifts. About 1/3 of the detected objects with good X-ray spectra are expected to be absorbed by column densities above $10^{22}$ cm$^{-2}$ (see Fig.\ref{zdist}~{\it right}). Future wide area optical and near-IR surveys should provide redshifts, either spectroscopic or photometric, for a significant fraction of these sources, making possible to divide the sample into fine redshift, luminosity and obscuration bins and therefore map the cosmological evolution of nuclear obscuration.

\begin{figure*}[t!]
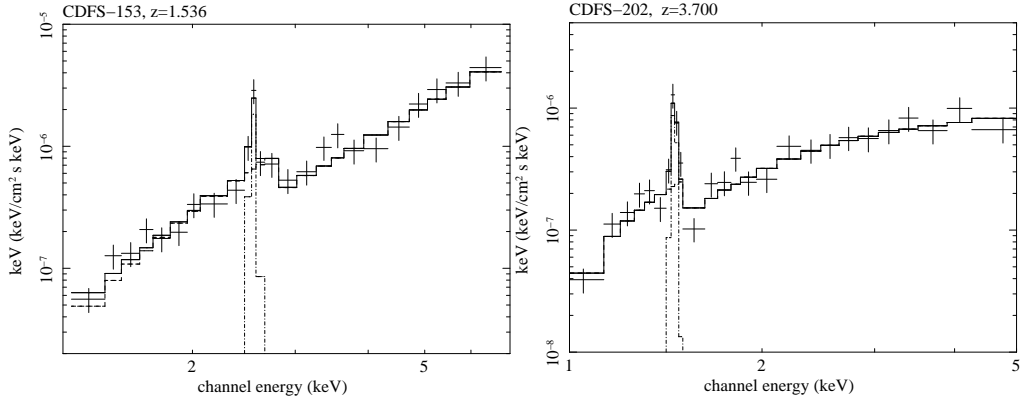

\resizebox{\hsize}{!}{
\includegraphics[angle=270]{cdfs153_deep_eeuf_xray2009.ps}
\includegraphics[angle=270]{cdfs202_deep_eeuf_xray2009.ps}}
\caption{\footnotesize Simulated WFXT spectra of CDFS-202 and CDFS-153, two high-$z$ CT AGN with $f_{2-10}>10^{-15}$ erg~cm$^{-2}$~s$^{-1}$ observed in the XMM-CDFS: the goal effective area and an exposure time of 400 ks were assumed. No background is included, which however is not expected to alter significantly the S/N ratio (see text).The 100 deg$^2$ WFXT-Deep survey \citep[Rosati et al. 2010, this volume]{murray08spie} is expected to reveal about 500 objects like these at $z>1$ in terms of obscuration and photon statistics.
}
\label{ct_simu}
\end{figure*}

\section{The most obscured AGN}

If the overall evolution of obscured AGN is uncertain, the evolution of the most heavily obscured and elusive ones, the so-called Compton-Thick AGN ($N_H>10^{24}$cm$^{-2}$; hereafter CT AGN), is completely unknown, making the census of accreting black holes largely incomplete.

Only $\sim$50 {\it bona-fide} CT AGN, i.e. certified by X-ray spectral analysis, are known in the local Universe \citep{c04,rdc08}, but their abundance has nonetheless been estimated to be comparable to that of less obscured ones \citep{guido99}. At higher redshifts, an integrated constrain to the space density of CT AGN can be obtained from the spectrum of the 
X-ray background. Depending on the assumptions, synthesis models of the XRB predict that from $\sim$10 to $\sim$30\% of the XRB peak emission at 30 keV is produced by the emission of CT AGN integrated over all redshifts \citep{gch07,tuv09}, but the main contribution likely arises at $z\sim1$. In the absence of any information, the luminosity function and evolution of CT AGN in XRB synthesis models have been usually assumed to be equal to those of less obscured ones. Because of the uncertainties
in the average spectrum of CT AGN and in their evolution, however, the constraints to the CT space density from the XRB spectrum remain rather loose.

In recent years there have been many attempts to constrain the space density of CT AGN in different luminosity and redshift intervals exploiting different selection techniques. Very hard ($>10$ keV) X-ray surveys are still limited in sensitivity and are just sampling the local Universe \citep{malizia09}. 
To sample AGN at higher redshifts, and in particular at $z\sim 1$, one needs to rely on deep X-ray surveys in the 2-10 keV band and try to select CT AGN either directly from X-ray spectroscopy, or by comparing the measured, obscured X-ray emission (if any) with some other indicator of the
intrinsic nuclear power: IR selection can track the nuclear emission as reprocessed by the dusty absorber; narrow optical emission lines can sample the gas ionized by the nucleus on scales free from obscuration. X-ray stacking of IR-selected sources not individually detected in the X-rays has been used to estimate the space density of CT AGN at $z \sim1-2$ \citep{daddi07,alex08,fiore09,bauer10}. The comparison between the [O~III]5007 and X-ray flux 
has been used to select X-ray underluminous QSOs and then estimate the density of CT QSOs at $z=0.5$ \citep{v10}. 
To sample the population of CT around $z\sim 1$, \citet{neon} recently devised a selection method based on the [Ne~V]3426 emission line, which can be applied to optical spectroscopic surveys with deep X-ray coverage and seems to deliver clean, albeit not complete, samples of CT objects at $z>0.8$, i.e. at redshifts not reachable with [O~III]5007 selection. Despite these numerous efforts, [Ne~V], [O~III], and IR selection are all indirect ways to select CT AGN, since the CT nature of an object is simply inferred from the faintness of its X-ray emission relative to an indicator of the intrinsic power, and therefore may suffer from severe systematic uncertainties.

To unambiguously obtain samples of  {\it bona-fide} CT AGN over a broad redshift range, very deep X-ray exposures are needed such as the 2.5 Ms XMM survey in the CDFS (Comastri et al. 2010, in prep.).  A number of CT candidates, identified in the 1 Ms CDFS catalog
on the basis of their flat (low quality) spectrum \citep{tozzi06}, are indeed being confirmed as such by the higher quality XMM spectra (Comastri et al. 2010, in prep.), including  the well-known CT candidate CDFS-202 at $z=3.7$ \citep{norman02}. Because of their limited sky coverage, however, only a few tens of {\it bona-fide} high-z CT AGN
are expected to be detected in current deep X-ray surveys, making population studies of CT AGN problematic.

The WFXT surveys are expected to determine the cosmological evolution of {\it bona-fide} CT AGN up to $z\sim 3$. Based on the synthesis model by \citet{gch07}, the WFXT-Deep survey will return a sample of $\sim 500$ objects at $z>1$ which are {\it bona-fide} CT AGN, i.e. with more than 500 net counts in the 0.5-7 keV band (the number of
simple detections of CT objects will be obviously much larger). In Fig.~\ref{ct_simu} we show two such objects (CDF-202 and CDFS-153, another CT AGN, at $z=1.53$, found in the XMM-CDFS)  simulated using the WFXT response matrices for the goal design. No background is assumed in the simulation, but, based on the
level estimated for low-earth orbits (see Ettori \& Molendi 2010, this volume), only about 20 background photons are expected above 1 keV, which would therefore
not alter significantly the spectral quality. It is evident that 500 X-ray photons are sufficient to unambiguously reveal their CT nature: in principle,
based on the iron K$\alpha$ line, it would be possible to determine also their redshift without the need of optical spectroscopy. In total, 500, 270, 60 and 12 {\it bona-fide} CT AGN at redshifts above 1, 2, 3 and 4, respectively, are expected in the WFXT-Deep survey. Future missions sensitive to energies above 10 keV, such as NuSTAR and ASTRO-H (approved by NASA and JAXA, respectively) or EXIST and NHXM (proposed to NASA and ASI, respectively) are expected to allow population studies of CT AGN at $z<1$, with a peak in the redshift distribution at $z\sim 0.3-0.4$. The WFXT mission appears to uniquely complement these and extend them to $z>1$. Only IXO, on a longer timescale and depending on the adopted survey strategy and mission design, could possibly provide samples of distant CT objects matching in size those expected from WFXT. 


\section{Conclusions}

About 15 millions of AGN in a very broad redshift and luminosity range will be detected by the WFXT surveys. Some of the major issues related
to the evolution of accreting SMBHs are expected to be solved by this sample as detailed below.
\\

\noindent
$\bullet$ WFXT will break through the high-z Universe: more than 1600 AGN will be observed at $z>6$, allowing population studies at these redshifts and providing an unvaluable complement for future wide area optical and near-IR surveys searching for black holes at the highest redshifts.
Such a large sample is a unique feature of the WFXT surveys and cannot be matched by any other planned or proposed X-ray mission.
\\

\noindent
$\bullet$ Good X-ray spectra, with more than 1000 photons, will be obtained for 300,000 AGN allowing accurate measurements of their absorbing column density. This will make possible for the first time to measure the evolution of nuclear obscuration with cosmic time up to $z\sim5$ and verify its connection with e.g. star formation.
\\

\noindent
$\bullet$ About 500 {\it bona fide}, i.e. certified by X-ray spectral analysis, CT AGN at $z>1$ will be found in the WFXT-Deep survey. This will complement population
studies at lower redshifts obtained by future high-energy surveys such as those performed by the approved missions NuSTAR and Astro-H
and will make possible to detemine the abundance and evolution of this still missing BH population. Performing population studies of distant heavily obscured 
objects, and determining their relevance in the census of accreting black holes and evolutionaly path of galaxies is another unique science case to be carried
out by WFXT. 

{
\acknowledgements

We thank all the members of the WFXT collaboration. RG acknowledges stimulating discussions with R. Salvaterra. We acknowledge partial support from ASI-INAF and PRIN/MIUR under grants I/023/05/00, I/088/06/00 and 2006-02-5203.

}

\bibliographystyle{aa}
\bibliography{../../neon/biblio} 

\end{document}